\documentclass[showpacs,amssymb,aps]{revtex4}

\newcommand{\be}{\begin{equation}}
\newcommand{\ee}{\end{equation}}

\usepackage{graphicx}

\begin{document}

\title{The static effective action for non-commutative QED at high
  temperature}  

\author{F. T. Brandt$^a$, Ashok Das$^b$, J. Frenkel$^a$,
S. Pereira$^a$ and  J. C. Taylor$^c$}
\affiliation{$^a$ Instituto de F\'{\i}sica,
Universidade de S\~ao Paulo,
S\~ao Paulo, SP 05315-970, BRAZIL}
\affiliation{$^b$Department of Physics and Astronomy,
University of Rochester,
Rochester, NY 14627-0171, USA}
\affiliation{$^c$ Department of Applied Mathematics and Theoretical
  Physics, University of Cambridge, Cambridge, UK}

\bigskip
%\date{}

\begin{abstract}

In this paper, we systematically study the effective action for
non-commutative QED in the static limit at high temperature. When
$\theta p^{2}\ll 1$, where $\theta$ represents the magnitude of the
parameter for non-commutativity and $p$ denotes a typical external
three momentum, we show that this leads naturally to
a derivative expansion in this model. The study of the self-energy,
in this limit, leads to nontrivial $\theta$ dependent corrections to
the electric and magnetic masses, which exist only above a certain
critical temperature. The three point and the four point amplitudes
are also studied as well as their relations to the Ward identities in
this limit. We determine the closed form
expression for the current involving only the spatial components of the
gauge field and present the corresponding
static effective action, which is gauge invariant.

\end{abstract}


\maketitle

\section{Introduction}

Thermal field theories \cite{kapusta:book89lebellac:book96das:book97}
are of  interest for a variety of
reasons. As is well known by now, thermal amplitudes and,
therefore, the effective actions have a non-analytic
structure \cite{weldon}. Consequently, they are best studied in some
limit. The static limit, where the external energies are set equal to
zero, is one such limit and is of interest in the study of a plasma at
very high temperatures because several physical quantities such as the
screening and the magnetic masses are defined in this limit. It is
also known that because of infrared divergences in a thermal field
theory, one needs to perform a resummation to obtain meaningful gauge
independent quantities at high temperature. While, in principle, the
resummation can involve general self-energy and vertex corrections (as
internal insertions),
the dominant contributions to the screening and magnetic masses come
from the static limit of these corrections (namely, the zero modes
contribute the most). It is for these reasons that the study of the
static limit of the effective action at high temperature is quite
useful. The hard thermal loops and the static effective actions in
conventional gauge theories have been well studied in the literature
\cite{Braaten:1990it,frenkel:1991ts}.

In this paper, we intend to carry out a corresponding analysis for
non-commutative QED. Non-commutative theories
\cite{Seiberg:1999vs,Fischler:2000fv,hayakawa,Arcioni:1999hw,
Landsteiner:2000bw,Szabo:2001kg,Douglas:2001ba,Chu:2001fe,VanRaamsdonk:2001jd,
Bonora:2000ga,Brandt:2002rw} are defined on a manifold where
coordinates do not commute, rather they satisfy
\begin{equation}
\left[x^{\mu},x^{\nu}\right] =
i\theta^{\mu\nu}\label{noncommutativity}
\end{equation}
where $\theta^{\mu\nu}$ is an anti-symmetric constant tensor. For
unitarity to hold in these theories \cite{gomis}, conventionally, one
assumes  that
$\theta^{0i}=0$, namely, we will assume that only the spatial
coordinates do not commute while the time coordinate commutes with
space coordinates. Furthermore, we note that the experimental
bound  on the magnitude of the parameter of non-commutativity leads to
\cite{sean} 
\begin{equation}
\theta = |\theta^{ij}| \leq (10\, {\rm TeV})^{-2} \approx
10^{-34}\,{\rm cm}^{2}\label{magnitude}
\end{equation}
The parameter for non-commutativity is, therefore, expected to be very
small. 

The non-commutativity of the coordinates leads to a modified product
on such a manifold, the Gr\"{o}enwald-Moyal star product, namely
\begin{equation}
f(x)\star g(x) =
e^{\frac{i}{2}\theta^{\mu\nu}\partial^{(\eta)}_{\mu}\partial^{(\xi)}_{\nu}}\,
\left. f(x+\eta)g(x+\xi)\right|_{\eta=\xi=0}\label{starproduct}
\end{equation}
As a consequence of the nontrivial nature of the star product (namely,
star products do not commute), the
Maxwell theory acquires a non-Abelian structure, namely, the action
for the Maxwell action on a non-commutative manifold takes the form
\begin{equation}
S = \int d^{4}x\,\left(-\frac{1}{4}\, F_{\mu\nu}\star
F^{\mu\nu}\right)\label{maxwellaction}
\end{equation}
where the field strength tensor has the form
\begin{equation}
F_{\mu\nu} = \partial_{\mu}A_{\nu} - \partial_{\nu}A_{\mu} -
ie\left[A_{\mu},A_{\nu}\right]_{\rm MB} = \partial_{\mu}A_{\nu} -
\partial_{\nu}A_{\mu} - ie \left(A_{\mu}\star A_{\nu} - A_{\nu}\star
A_{\mu}\right)\label{fieldstrength}
\end{equation}
The action (\ref{maxwellaction}) is invariant under a gauge
transformation
\begin{equation}
A_{\mu} \rightarrow U\star A_{\mu}\star U^{-1} - \frac{i}{e}\,U\star
\partial_{\mu}U^{-1}\label{gauge transformation}
\end{equation}
which is reminiscent of non-Abelian gauge transformations in
conventional theories. The structure of the field strength tensor in
(\ref{fieldstrength}) also makes it clear that Maxwell's theory on a
non-commutative manifold involves self-interactions. Consequently,
since the action in (\ref{maxwellaction}) is an interacting theory, we
neglect the fermions, although we can add fermions in a natural
manner. There is a second reason for neglecting the fermions. It is
known that fermion loops only lead to planar contributions which are
the same as in  conventional QED and we are interested in $\theta$
dependent corrections to various physical quantities.

The paper is organized as follows. In section {\bf II}, we describe in
detail the tensor structure for the self-energy in non-commutative QED
at finite temperature. We also give the perturbative result for the
self-energy in the static limit. This can be exactly evaluated in a
closed form, as was observed earlier \cite{Brandt:2002aa}. Here, we
clarify  the reason for
such a simplification. We determine the $\theta$ dependent screening
and the  magnetic masses in this theory at the one loop level and show
that these contributions are nontrivial only for temperatures above a
certain temperature.  In
section {\bf III}, we study the leading terms in the three point and
the four  point
amplitudes in some detail and show that their structure is consistent
with what we will expect from the Ward identities. In fact, the three
point function can be completely expressed in terms of the static
self-energy. This is a consequence of the fact that amplitudes with an
odd number of temporal indices (such as $\Gamma_{000}$) vanish. On the
other hand, not all nontrivial components of the four point function
can be expressed in terms of
the lower order amplitudes, since, in this case, $\Gamma_{0000}$ neither
vanishes nor is constrained by the Ward identity and, consequently,
needs to be evaluated independently. In section
{\bf IV}, we solve the Ward identity and determine, in terms of the
self-energy, a simple
expression for the current which depends on the spatial components of
the gauge field. 
In section {\bf V}, we present a closed form effective action for the
static amplitudes, with spatial tensor structures, which is valid at
high temperatures in the region $\theta p^2\ll 1$. This gauge
invariant action (see Eq. (\ref{jct5})) is expressed in terms of
functions which may be related to open Wilson lines.

\section{Self-energy for non-commutative QED in the static limit at
  high temperature} 

In this section, we will discuss the tensor decomposition of the
self-energy in non-commutative QED at finite temperature. Using this,
we will  evaluate the
self-energy in the static limit at high temperature and study various
masses that follow.

Let us begin by recalling that in a conventional theory, at zero
temperature,  there are two
natural tensor structures, $\eta^{\mu\nu}$ and $p^{\mu}$, the
external momentum, with which we can describe the self-energy. In a
non-commutative theory at finite temperature, we have additional
structures such as $\theta^{\mu\nu}$ and $u^{\mu}$, the
velocity of the heat bath. To determine the most general, second rank
symmetric tensor constructed from $\eta^{\mu\nu},
p^{\mu},\theta^{\mu\nu},u^{\mu}$, let us proceed as follows. First, we
note that there are seven distinct second rank symmetric tensor
structures that we can form, namely, $\eta^{\mu\nu},
u^{\mu}u^{\nu},p^{\mu}p^{\nu},
\tilde{p}^{\mu}\tilde{p}^{\nu},(p^{\mu}u^{\nu}+p^{\nu}u^{\mu}),
(p^{\mu}\tilde{p}^{\nu}+p^{\nu}\tilde{p}^{\mu}),
(\tilde{p}^{\mu}u^{\nu}+\tilde{p}^{\nu}u^{\mu})$ where we have defined
\begin{equation}
\tilde{p}^{\mu} = \theta^{\mu\nu}p_{\nu}\label{ptilde}
\end{equation}
By definition, $\tilde{p}^{\mu}$ is transverse to $p^{\mu}$ and,
furthermore, it can also be easily verified that $u\cdot \tilde{p}=0$
since $\theta^{\mu\nu}$ only involves spatial indices. However, to
leading order at high temperature, the Ward identities require that
the self-energy be transverse to the external momentum. To obtain the
most general second rank symmetric tensor that is also transverse, let
us define  
\begin{eqnarray}
\hat{\eta}^{\mu\nu} & = & \eta^{\mu\nu} -
u^{\mu}u^{\nu}\nonumber\\
\noalign{\vskip 4pt}%
\hat{p}^{\mu} & = & p^{\mu} - (u\cdot p) u^{\mu}\nonumber\\
\noalign{\vskip 4pt}%
\bar{u}^{\mu} & = & u^{\mu} - \frac{(u\cdot
  p)}{p^{2}}\,p^{\mu}\label{transversetensors}
\end{eqnarray}
By construction, the ``hat'' variables are orthogonal to $u^{\mu}$
(the velocity is normalized to unity, $u\cdot u=1$) while
$\bar{u}^{\mu}$ is orthogonal to $p^{\mu}$. It is
easy to see now that we can construct four independent second rank symmetric
tensors which are transverse so that the self-energy, for the photon,
can be  written in the form
\begin{equation}
\Pi^{\mu\nu} =
A\,\left(\hat{\eta}^{\mu\nu}-\frac{\hat{p}^{\mu}\hat{p}^{\nu}}{\hat{p}^{2}}
\right) + B\, \frac{p^{2}}{\hat{p}^{2}}\,\bar{u}^{\mu}\bar{u}^{\nu} + 
C\,\frac{\tilde{p}^{\mu}\tilde{p}^{\nu}}{\tilde{p}^{2}}
+D\,(\bar{u}^{\mu}\tilde{p}^{\nu}+\bar{u}^{\nu}\tilde{p}^{\mu})\label{general} 
\end{equation}
However, we note that the self-energy for the photon is even under
charge conjugation ($\theta\rightarrow -\theta$)
\cite{Sheikh-Jabbari:1999vm,fernando}, while the last
structure in (\ref{general}) is odd. Therefore, we must have $D=0$ and
to all  orders, the
self energy can be parameterized as
\begin{equation}
\Pi^{\mu\nu} = A\,P^{\mu\nu} + B\,Q^{\mu\nu} +
C\,R^{\mu\nu}\label{decomposition}
\end{equation}
where we have defined
\begin{equation}
P^{\mu\nu} = \left(\hat{\eta}^{\mu\nu} -
\frac{\hat{p}^{\mu}\hat{p}^{\nu}}{\hat{p}^{2}}\right),\quad Q^{\mu\nu}
= \frac{p^{2}}{\hat{p}^{2}}\,\bar{u}^{\mu}\bar{u}^{\nu},\quad
R^{\mu\nu} =
\frac{\tilde{p}^{\mu}\tilde{p}^{\nu}}{\tilde{p}^{2}}\label{projection}
\end{equation}

The tensors appearing in (\ref{projection}) are easily seen to be
projection operators,
\begin{equation}
P^{\mu\lambda}P_{\lambda\nu} = P^{\mu}_{\nu},\quad
Q^{\mu\lambda}Q_{\lambda\nu} = Q^{\mu}_{\nu},\quad
R^{\mu\lambda}R_{\lambda\nu} = R^{\mu}_{\nu}
\end{equation}
However, they are not orthonormal. In fact, it is easy to check that
\begin{equation}
P^{\mu\lambda}Q_{\lambda\nu} = 0 = Q^{\mu\lambda}R_{\lambda\nu},\quad
P^{\mu\lambda}R_{\lambda\nu} = R^{\mu}_{\nu}\label{orthonormality}
\end{equation}
This suggests that a better basis to work with is given by
$\overline{P}^{\mu\nu},Q^{\mu\nu},R^{\mu\nu}$ where
\begin{equation}
\overline{P}^{\mu\nu} = P^{\mu\nu} - R^{\mu\nu}\label{pbar}
\end{equation}
so that all the structures correspond to orthonormal projection operators. In
this basis, we can parameterize the leading order self-energy at high
temperature  as
\begin{equation}
\Pi^{\mu\nu} = \overline{P}^{\mu\nu}\,\Pi_{\rm T} +
Q^{\mu\nu}\,\Pi_{\rm L} +
R^{\mu\nu}\,\tilde{\Pi}_{T}\label{tensordecomposition}
\end{equation}
The meaning of the various projections is quite clear. While
$\overline{P}^{\mu\nu},Q^{\mu\nu},R^{\mu\nu}$ are all orthogonal to
$p^{\mu}$, it is easy to see from their definitions in
(\ref{projection}) and (\ref{pbar}) that $\overline{P}^{\mu\nu}$ is, in
addition, orthogonal to $u^{\mu}$ as well as to
$\tilde{p}^{\mu}$. Similarly, $Q^{\mu\nu}$ is additionally transverse
to $\tilde{p}^{\mu}$ and $R^{\mu\nu}$ to $u^{\mu}$. Thus,
additionally, $\overline{P}^{\mu\nu}$ and $R^{\mu\nu}$ are transverse
to $p^{i}$ (that is the reason for the subscript ``T'' in their form
factors) while $Q^{\mu\nu}$ is not (which is why the subscript on the
form factor is ``L''). Furthermore, while $\overline{P}^{\mu\nu}$ and
$R^{\mu\nu}$ are both orthogonal to $p^{i}$, the first is orthogonal
to vectors in the non-commutative plane (if only two coordinates do
not commute) while the second is not. Finally, let us note that
\begin{equation}
\overline{P}^{\mu\nu} + Q^{\mu\nu} + R^{\mu\nu} = \eta^{\mu\nu} -
\frac{p^{\mu}p^{\nu}}{p^{2}}\label{sum}
\end{equation}

With the parameterization of the self-energy in
(\ref{tensordecomposition}) in terms of orthonormal projection
operators, several things simplify. First, we note that we can
determine the various form factors as
\begin{equation}
\Pi_{\rm L} =
\frac{p^{2}}{\hat{p}^{2}}\,u_{\mu}u_{\nu}\Pi^{\mu\nu},\quad
\tilde{\Pi}_{\rm T} =
\frac{\tilde{p}_{\mu}\tilde{p}_{\nu}}{\tilde{p}^{2}}\,\Pi^{\mu\nu},\quad
(D-3)\Pi_{\rm T} = \eta_{\mu\nu}\Pi^{\mu\nu} - \Pi_{\rm L} -
\tilde{\Pi}_{\rm T}\label{formfactors}
\end{equation}
Here, $D$ represents the number of space-time dimensions. In
particular, we note that when $D=3$, we do not have any information on
the transverse form factor from these equations which has to be
contrasted with the case in a  conventional theory (for which the same
happens if $D=2$). Adding in the tree level two point function, we can
write to all orders
\begin{equation}
\Gamma^{\mu\nu} = \overline{P}^{\mu\nu}\,(p^{2}+\Pi_{\rm T}) +
Q^{\mu\nu}\,(p^{2}+\Pi_{\rm L}) + R^{\mu\nu}\,(p^{2}+\tilde{\Pi}_{\rm
  T}) + \frac{p_{\mu}p_{\nu}}{\xi}\label{twopoint}
\end{equation}
where $\xi$ represents the gauge fixing parameter in a covariant
gauge. Since the projection operators are orthonormal, the inverse can
be easily obtained, leading to the propagator
\begin{equation}
D_{\mu\nu} = \overline{P}_{\mu\nu}\,\frac{1}{p^{2}+\Pi_{\rm T}} +
Q_{\mu\nu}\,\frac{1}{p^{2}+\Pi_{\rm L}} +
R_{\mu\nu}\,\frac{1}{p^{2}+\tilde{\Pi}_{\rm T}} +
\xi\,\frac{p_{\mu}p_{\nu}}{p^{2}}\label{propagator}
\end{equation}
The poles in the propagator are distinct as a consequence of our
choice of orthonormal projection operators (Had we used a different
basis, the poles will be mixed and will need to be disentangled). We
see that there are three physical poles (in addition to the unphysical
one coming from the gauge fixing). The meaning of the three poles is
easily understood as follows. First, we can define the screening mass,
as in a conventional theory, as (our Minkowski metric has the
signatures $(+,-,-,-)$)
\begin{equation}
m_{\rm el}^{2} = - \Pi_{\rm L} (p_{0}=0, p^{2} = -{\bf p}^{2} =  m_{\rm
  el}^{2})\label{screening}
\end{equation}
The conventional magnetic mass can also be defined as
\begin{equation}
m_{\rm mag}^{2} = - \Pi_{\rm T} (p_{0}=0,p^{2} = -{\bf p}^{2} =  m_{\rm
  mag}^{2})\label{magnetic}
\end{equation}
However, there is now a new transverse pole at
\begin{equation}
\tilde{m}_{\rm mag}^{2} = - \tilde{\Pi}_{\rm T} (p_{0}=0,p^{2}=-{\bf
  p}^{2} = \tilde{m}_{\rm mag}^{2})\label{newmagnetic}
\end{equation}
This can be thought of as the screening length between magnetic fields
in the non-commutative plane. This feature is new in  non-commutative
QED, since the non-commutative parameter can define a preferred
direction in space.

\begin{figure}[h!]
% The following should be used when creating pdf directly out of
% latex, using the command line pdflatex
%    \includegraphics[scale=1.0]{two_photon.pdf}
\includegraphics[scale=1.0]{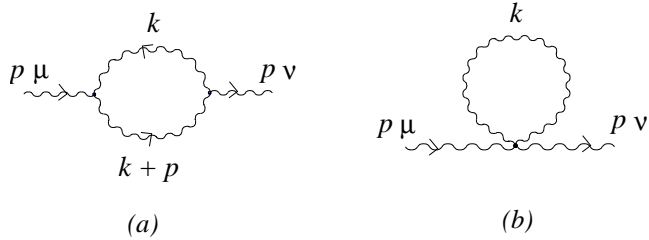}
\caption{One-loop diagrams which contribute to the 
photon self-energy in non-commutative QED. The wavy lines
represent photons and diagrams with ghost loops are understood to be
included.}\label{fig0} 
  \end{figure}

Let us now evaluate the self-energy, represented in figure 1, in
the static limit at high temperature. We note that the calculation of
the self-energy, in the static limit, was already done in
\cite{Brandt:2002aa}  and the
result was surprisingly very simple. Here, we would like to understand
the reason for the simplicity of this result and then calculate the
physical masses in the theory. To begin with, let us tabulate a
few integrals \cite{gradshteyn} that will be useful in the evaluation of the
self-energy.
\begin{eqnarray}
\int_{0}^{\infty} dx\,\frac{x}{e^{\frac{x}{T}}-1} & = &
\frac{\pi^{2}T^{2}}{6}\nonumber\\
\noalign{\vskip 4pt}%
\int_{0}^{\infty} dx\,\frac{\sin xy}{e^{\frac{x}{T}}-1} & = & 
\frac{\pi T}{2}\left(\coth \pi yT - \frac{1}{\pi yT}\right)\nonumber\\
\noalign{\vskip 4pt}%
\int_{0}^{\infty} dx\,\frac{x\cos xy}{e^{\frac{x}{T}}-1} & = &
\frac{1}{2y^{2}} - \frac{\pi^{2}T^{2}}{2}\,{\rm cosech}^{2} \pi
yT\label{integrals}
\end{eqnarray}
A direct application of the forward scattering amplitude method
\cite{brandt:1993mj,brandt:1993dkbrandt:1997se} leads, in the hard
thermal loop approximation, 
to the the self-energy of the form
\begin{equation}
\Pi^{\mu\nu} (p) = - \frac{4e^{2}}{(2\pi)^{3}} \int
d^{3}k\,\frac{n_{B}(k)}{k} (1 - \cos \tilde{p}\cdot
k)\left.\left[\eta^{\mu\nu} -
  \frac{p^{\mu}k^{\nu}+p^{\nu}k^{\mu}}{p\cdot k} +
  \frac{p^{2}k^{\mu}k^{\nu}}{(p\cdot
    k)^{2}}\right]\right|_{k^{0}=k}\label{barton}
\end{equation}
where $k = |{\bf k}|$ and $n_{B}$ represents the bosonic distribution
function. Let us recall that the hard thermal loop approximation, in
this theory, involves assuming
\begin{equation}
p\ll k \sim {\rm min}\,(T, \frac{1}{\tilde{p}})\label{hardthermalloop}
\end{equation}
Going to the rest frame of the heat bath and using
(\ref{integrals}), it now follows easily that
\begin{eqnarray}
\eta_{\mu\nu}\Pi^{\mu\nu} & = & - \frac{8e^{2}}{(2\pi)^{3}} \int
d^{3}k\,\frac{n_{B}(k)}{k} (1 - \cos \tilde{p}\cdot k)\nonumber\\
\noalign{\vskip 4pt}%
 & = & - \frac{16e^{2}}{(2\pi)^{2}} \int_{0}^{\infty}
\frac{dk\,k}{e^{\frac{k}{T}}-1}\,\left(1 - \frac{\sin
  k|\tilde{p}|}{k|\tilde{p}|}\right) = - 2e^{2}T^{2}\left[\frac{1}{3}
  - \frac{1}{\pi |\tilde{p}|T}\left(\coth \pi |\tilde{p}|T -
  \frac{1}{\pi |\tilde{p}|T}\right)\right]\label{trace}
\end{eqnarray}
where we have defined
\begin{equation}
|\tilde{p}| = |\theta^{ij}p_{j}|\label{norm}
\end{equation}

While the calculation of the trace of the self-energy from
(\ref{barton}) is simple, in the static limit, the calculations of
$\Pi_{\rm L},\tilde{\Pi}_{\rm T}$ are not, and are manifestly 
non-local. However, with a little bit of algebra, which involves
integration by parts of the relation
\begin{equation}
\frac{p^{\mu}k^{\nu}+p^{\nu}k^{\mu}}{p\cdot k} -
\frac{p^{2}k^{\mu}k^{\nu}}{(p\cdot k)^{2}} =
p_{\lambda}\frac{\partial}{\partial k_{\lambda}}
\left(\frac{k^{\mu}k^{\nu}}{p\cdot k}\right)
\end{equation}
it may be shown that Eq. (\ref{barton}) can be rewritten as
\begin{equation}
\Pi^{\mu\nu} =  - \frac{4e^{2}}{(2\pi)^{3}} \int
\frac{d^{3}k}{k}\,(1-\cos \tilde{p}\cdot
k)\left.\left[\eta^{\mu\nu} n_{B}(k)
  +n'_{B}(k)\,\frac{p^{0}k^{\mu}k^{\nu}}{p\cdot k} - (kn'_{B}(k) 
  - n_{B}(k)) \frac{k^{\mu}k^{\nu}}{k^{2}} - n_{B}(k) \frac{\eta^{\mu
      0}k^{\nu}+\eta^{\nu
      0}k^{\mu}}{k}\right]\right|_{k^{0}=k}\label{integrated} 
\end{equation}
where a prime denotes differentiation with respect to $k$. It is clear
from  (\ref{integrated}) that the potentially non-local
terms vanish in the static limit when $p^{0}=0$. Thus, we see that the
self-energy is a local function in the static limit, with a simple
form (obtained by using the symmetry of the ${\bf k}$ integral)
\begin{equation}
\Pi^{\mu\nu}_{\rm static} =  \frac{4e^{2}}{(2\pi)^{3}} \int
\frac{d^{3}k}{k}\,(1-\cos\tilde{p}\cdot k)\left.\left[-\eta^{\mu\nu}
  n_{B}(k) + (kn'_{B}(k) - n_{B}(k)) \frac{k^{\mu}k^{\nu}}{k^{2}} +
  n_{B}(k) \frac{2\eta^{\mu 0}\eta^{\nu
      0}}{k}\right]\right|_{k^{0}=k}\label{static}
\end{equation}
There are several things to note from (\ref{static}). First, the
integrand, except for the trigonometric function (coming from the
vertices of the non-commutative theory), is completely local and is
independent of the external momentum. Since the trigonometric function
does not involve $k^{0}$ (namely, $\theta^{0i}=0$), it can be taken
outside the Matsubara sum in the imaginary time formalism and it
is clear  that the result, (\ref{static}), can be obtained directly
from the  Matsubara sum of
frequencies by setting the external momentum equal to zero (except in
the trigonometric factor which is outside the sum and will give zero
if the external momentum is naively set to zero). In this
case, the  sum is very simple and can be done in
a trivial manner. In this sense, this result can be understood as the
leading term in a derivative expansion. This is, in fact, supported by
the structure of the theory. We know that amplitudes become
non-analytic in a thermal field theory. However, once we are in the
static limit, the amplitudes are analytic in $p^{i}$ (in the absence
of  infrared problems) so that a derivative expansion does make
sense. We have shown earlier that although the amplitudes in a
non-commutative theory are also non-analytic, the non-analyticity is
not a consequence of any new branch cut. Therefore, we expect the
general analytic behavior of the conventional thermal field theories
to hold in a non-commutative theory at finite
temperature. Furthermore, we note that because of the trigonometric
function in (\ref{static}), in the infrared limit $(1-\cos
\tilde{p}\cdot k)\rightarrow 0$ and, consequently, infrared divergence
is not a problem in such theories at finite temperature (namely, as
$p^{i}\rightarrow 0$, the coupling vanishes in such
theories). Therefore, in
the static limit, we expect the amplitudes to be analytic in $p^{i}$,
leading to the fact that a derivative expansion can be carried
out. This also explains the simplicity of the form for the self-energy
in the static limit. Namely, if we set all the external momentum to
zero in the denominator (namely, the leading term in the derivative
expansion), then, the integrand involves only one angular integral
coming from the trigonometric function which is easy to carry out. We
also note from the form of the amplitude in (\ref{static}) that
$\Pi^{0i} = \Pi^{i0}=0$ from the symmetry of the integrand. We will
comment more on this in the next section. 

The components of the self-energy, in the static limit, can now be
easily calculated. 
Without going into the details, we simply note that, in the rest frame
of the  heat bath, the components of the self-energy take the forms
\begin{eqnarray}
\Pi^{00}_{\rm static} & = & - \frac{2e^{2}T^{2}}{3}\left[1 -
  \frac{3}{2}\left(\frac{\coth \pi |\tilde{p}|T}{\pi |\tilde{p}|T} -
  {\rm cosech}^{2} \pi |\tilde{p}|T\right)\right]\nonumber\\
\noalign{\vskip 4pt}%
\Pi^{0i}_{\rm static} & = & 0\nonumber\\
\noalign{\vskip 4pt}%
\Pi^{ij}_{\rm static} & = & - e^{2}T^{2}\left[\frac{\coth \pi |\tilde{p}|T}{\pi
  |\tilde{p}|T} + {\rm cosech}^{2} \pi |\tilde{p}|T - \frac{2}{(\pi
  |\tilde{p}|T)^{2}}\right]
  \frac{\tilde{p}^{i}\tilde{p}^{j}}{\tilde{p}^{2}}\label{components}
\end{eqnarray}
Therefore, in this case, we have (see (\ref{formfactors}) in the rest
frame of the heat bath)
\begin{eqnarray}
\Pi_{\rm L}^{\rm static} & = & \Pi^{00}_{\rm static} =  -
  \frac{2e^{2}T^{2}}{3} \left[1 -
  \frac{3}{2}\left(\frac{\coth \pi |\tilde{p}|T}{\pi |\tilde{p}|T} -
  {\rm cosech}^{2} \pi |\tilde{p}|T\right)\right]\nonumber\\
\noalign{\vskip 4pt}%
\tilde{\Pi}_{\rm T}^{\rm static} & = &
  \frac{\tilde{p}_{i}\tilde{p}_{j}}{\tilde{p}^{2}} \Pi^{ij}_{\rm static} = -
  e^{2}T^{2}\left[\frac{\coth \pi |\tilde{p}|T}{\pi 
  |\tilde{p}|T} + {\rm cosech}^{2} \pi |\tilde{p}|T - \frac{2}{(\pi
  |\tilde{p}|T)^{2}}\right]\nonumber\\
\noalign{\vskip 4pt}%
\Pi_{\rm T}^{\rm static} & = & \eta_{\mu\nu} \Pi^{\mu\nu}_{\rm static} -
  \Pi_{\rm L}^{\rm static} - \tilde{\Pi}_{\rm
  T}^{\rm static} = 0\label{formfactors1}
\end{eqnarray}
This shows that the conventional magnetic mass, $m_{\rm mag}$ defined
in (\ref{magnetic}), vanishes as in QED on a commutative manifold. In
the static limit, therefore, the self-energy 
(\ref{tensordecomposition}) takes the form
\begin{equation}
\Pi^{\mu\nu}_{\rm static} = u^{\mu}u^{\nu}\,\Pi_{\rm L}^{\rm static} +
\frac{\tilde{p}^{\mu}\tilde{p}^{\nu}}{\tilde{p}^{2}}\,\tilde{\Pi}_{\rm
  T}^{\rm static}\label{static1}
\end{equation}

On the other hand, we see that both $\Pi_{\rm L}^{\rm
  static},\tilde{\Pi}_{\rm T}^{\rm static}$
have nontrivial contributions depending on $\theta$ (through
$\tilde{p}$). This is to be expected since the effect of
non-commutativity can be classically thought of as being equivalent to
a background electromagnetic field. We note, in particular, that since
$\tilde{\Pi}_{\rm T}^{\rm static}$ is nontrivial, there is a
  possibility, in  this
theory, to have a nontrivial magnetic mass in the non-commutative
plane, even though the conventional magnetic mass vanishes. The
screening mass and the ``new'' magnetic mass can be determined from
the equations (see (\ref{screening}) and (\ref{newmagnetic}))
\begin{eqnarray}
m_{\rm el}^{2} & = & - \Pi_{\rm L}^{\rm static} ({\bf p}^{2}=-m_{\rm
  el}^{2})\nonumber\\
\noalign{\vskip 4pt}%
\tilde{m}_{\rm mag}^{2} & = & - \tilde{\Pi}_{\rm T}^{\rm static} ({\bf
  p}^{2}=-\tilde{m}_{\rm mag}^{2})\label{masses}
\end{eqnarray}
\begin{figure}[h!]
% The following should be used when creating pdf directly out of
% latex, using the command line pdflatex
%    \includegraphics[scale=0.5]{masses.pdf}
\includegraphics[scale=0.5]{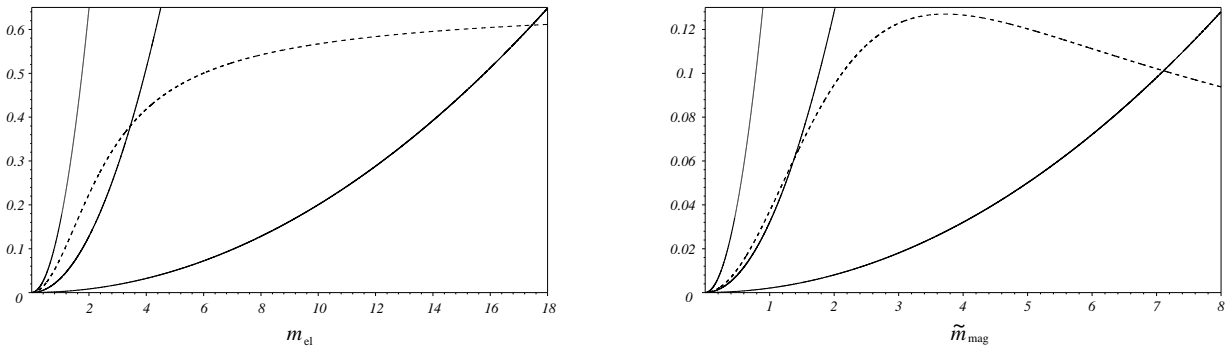}
\caption{The electric and the magnetic masses in units of 
$(\pi\,\theta\,T)^{-1}$.
The three solid lines are the plots of the parabola
corresponding to the left hand side of Eqs. (\ref{masses}) for
$ (e\,\pi\,\theta\,T^2) =2.5,\;5.6\;{\rm and}\; 22.4$.
The corresponding right hand sides are plotted using dashed lines.}\label{fig1}
  \end{figure}
These simultaneous equations can be solved graphically (see
figure \ref{fig1}). We choose a coordinate system in which
$\theta_{12}=-\theta_{21} = \theta$ represent the only non-vanishing
components of $\theta_{ij}$. Then, setting $p^{3}=0$, 
we note that, in both the equations, the left and the right
hand side give rise to parabola near the origin and, consequently,
unless the slopes have appropriate values, there will be no
intersection of  the
curves (and, therefore, no solution). This leads to the fact that,
for a nontrivial screening mass to exist in this theory, we must have
\begin{equation}
T^{2} > T^{2}_{\rm c} =
\frac{3\sqrt{5}}{2\pi e\theta}\label{tcscreeninbg}
\end{equation}
Similarly, for a nontrivial ``new'' magnetic mass to exist, we must
have
\begin{equation}
T^{2} > T_{\rm c}^{2} = \frac{3\sqrt{10}}{2\pi
    e\theta}\label{tcmagnetic}
\end{equation}
This is very interesting in that such a mass develops only above a
critical temperature. Considering the smallness of $\theta$ (see
(\ref{magnitude})), we recognize that these temperatures are very
high. Nonetheless, as a matter of principle, it is interesting to note
that this behavior is quite similar to the propagation of waves in a
wave guide or a plasma, which exists only above a critical cut-off
frequency.

\section{Higher point amplitudes in the static limit at high
  temperature}

In studying the higher point functions, in the static limit, at high
temperature, we note that the complete symmetry of the amplitudes in the
leading  order approximation of the derivative expansion, leads to
the  result that any amplitude with
an odd number of temporal indices vanishes. This is
already evident in the results of the last section, namely,
$\Pi^{0i}=0$. Therefore, we can concentrate only on amplitudes with an even
number of temporal indices. In the case of the three point amplitude,
this implies that we must have
\begin{equation}
\Gamma^{000}_{\rm static} = 0 = \Gamma^{0ij}_{\rm static}
\end{equation}
and the only nontrivial components of the three point amplitude can be
identified with $\Gamma^{00i},\Gamma^{ijk}$. Explicit calculations
bear out this expectation.

\begin{figure}[h!]
% The following should be used when creating pdf directly out of
% latex, using the command line pdflatex
%    \includegraphics[scale=0.8]{three_photon.pdf}
\includegraphics[scale=0.8]{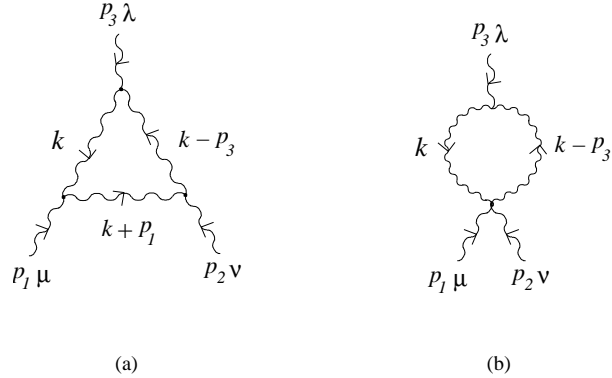}
\caption{Typical one-loop diagrams for the three point photon amplitude in
  non-commutative QED.}\label{fig2}
  \end{figure}

From the discussion of the last section, we note that the leading
contributions to any amplitude, in the static limit, can be obtained
from the lowest order terms in a derivative expansion. Such a derivative
expansion, as we have seen, corresponds to setting the external
momenta equal to zero everywhere in the integrand except in the
trigonometric functions. We note that the terms in the integrand, other
than the trigonometric functions, have the general behavior that, in
the hard thermal loop approximation, they are functions of zero degree in
the external four momenta. Therefore, in the static limit, these
factors  become
independent of the spatial momenta giving rise to the appearance of
the leading contribution in a particular derivative expansion. The
trigonometric functions, on the other hand, do not have this property.
In the trigonometric functions, however, we
can neglect contributions quadratic in the external momenta compared
to terms linear in the external momenta. Thus, for example, in the
three point amplitude diagram coming from three cubic vertices
(see figure \ref{fig2}-a), the trigonometric functions
coming from  the vertices, can be simplified as
\begin{equation}
\sin \left(\frac{\tilde{p}_{1}\cdot k}{2}\right)\sin
\left(\frac{\tilde{p}_{2}\cdot (k-p_{3})}{2}\right) \sin
\left(\frac{\tilde{p}_{3}\cdot k}{2}\right) \approx \sin
\left(\frac{\tilde{p}_{1}\cdot k}{2}\right) \sin
\left(\frac{\tilde{p}_{2}\cdot k}{2}\right) \sin
\left(\frac{\tilde{p}_{3}\cdot k}{2}\right)\label{simplification}
\end{equation}
Expanding the second trigonometric function on the left hand side, it
is easy to see that this corresponds to using the approximation that
\begin{equation}
\theta p^{2}\ll 1\label{approx}
\end{equation}
where $p$ denotes the typical magnitude of the
external momentum. Mathematically, such a derivative expansion would
correspond to choosing
\begin{equation}
p\ll k, \qquad \theta pT\sim O(1)
\end{equation}
which would automatically lead to (\ref{approx}).

Since the trigonometric functions do not involve
any dependence on the energy ($\theta^{0i}=0$), in the regime
(\ref{approx}), the  calculation of any
higher point amplitude, in the static limit, simplifies enormously and
can be carried out directly in the imaginary time formalism. Explicit
calculations show that, when all the graphs contributing to a given
amplitude are summed, the trigonometric functions in the integrand of
the $n$-point amplitude correspond to a product of $n$ factors of
$\sin \left(\frac{\tilde{p}_{i}\cdot k}{2}\right)$ with $i=1,2,\cdots
n$. This is consistent with the symmetry expected of the total amplitude,
namely, since the only dependence on the external momenta is in the
trigonometric 
functions in the leading order, and since the amplitude has to be symmetric
under the exchange of external bosonic lines, the trigonometric
functions must reflect this also. However, it is worth noting here
that this is not expected to hold for individual graphs which is
evident in the explicit calculations.

The recipe for calculating any higher point amplitude is now
clear. For the $n$-point amplitude, for example, the integrand will
involve $n$ trigonometric factors which can be taken outside the
Matsubara sum, which has no dependence on the external momentum. Thus, for
the  three point amplitude, we obtain
\begin{equation}
\Gamma_{\mu\nu\lambda}^{\rm static} = ie^{3}T \int
\frac{d^{3}k}{(2\pi)^{3}}\, \sin
\left(\frac{\tilde{p}_{1}\cdot k}{2}\right)\sin
\left(\frac{\tilde{p}_{2}\cdot k}{2}\right) \sin
\left(\frac{\tilde{p}_{3}\cdot k}{2}\right) \sum_{n}
\left[\frac{128 k_{\mu}k_{\nu}k_{\lambda}}{((2\pi nT)^{2}+k^{2})^{3}} -
  \left(\frac{32 \delta_{\mu\nu} k_{\lambda}}{((2\pi
    nT)^{2}+k^{2})^{2}}+ {\rm cyclic}\right)\right]\label{threepoint}
\end{equation}
Although (\ref{threepoint}) appears to
involve three angles coming from the trigonometric functions (in which
case the integration over spatial components would be nontrivial), we
can use the identity
\begin{equation}
\sin\left(\frac{\tilde{p}_{1}\cdot
  k}{2}\right)\sin\left(\frac{\tilde{p}_{2}\cdot
  k}{2}\right)\sin\left(\frac{\tilde{p}_{3}\cdot k}{2}\right) = -
  \frac{1}{4}\left(\sin \tilde{p}_{1}\cdot k + \sin \tilde{p}_{2}\cdot
  k + \sin \tilde{p}_{3}\cdot k\right)
\end{equation}
This is nice since each term involves only one angular integral which
can be carried out using (\ref{integrals}). Then, (\ref{threepoint})
becomes 
\begin{equation}
\Gamma_{\mu\nu\lambda}^{\rm static} = -8ie^{3}T \int
\frac{d^{3}k}{(2\pi)^{3}}\left( \sin \tilde{p}_{1}\cdot k + \sin
\tilde{p}_{2}\cdot k + \sin \tilde{p}_{3}\cdot k \right) \sum_{n}
\left[\frac{4k_{\mu}k_{\nu}k_{\lambda}}{((2\pi nT)^{2}+k^{2})^{3}} -
  \left(\frac{\delta_{\mu\nu} k_{\lambda}}{((2\pi
    nT)^{2}+k^{2})^{2}}+ {\rm cyclic}\right)\right]\label{threepoint1}
\end{equation}
It is worth noting from this expression that when there is an odd
number of temporal indices, the amplitude vanishes because of
anti-symmetry in the Matsubara sum, which is consistent with the
general structure of the static amplitudes in the leading order.

The actual evaluation of the thermal parts from the Matsubara sums can
be  carried out using the following relations 
\begin{eqnarray}
T \sum_{n} \frac{1}{(2\pi nT)^{2}+ k^{2}} & = & \frac{n_{B}(k)}{k} +
(T=0\,{\rm term})\nonumber\\
\noalign{\vskip 4pt}%
T \sum_{n} \frac{1}{((2\pi nT)^{2} + k^{2})^{2}} & = & -
\frac{1}{2k}\,\left(\frac{n_{B}(k)}{k}\right)' + (T=0\,{\rm
  term})\nonumber\\
\noalign{\vskip 4pt}%
T \sum_{n} \frac{1}{((2\pi nT)^{2} + k^{2})^{3}} & = &
\frac{1}{4k}\left[\frac{1}{2k}\left(\frac{n_{B}(k)}{k}\right)'\right]'
+ (T=0\,{\rm term})\label{sums}
\end{eqnarray}
where prime denotes a derivative with respect to $k$. Using these as
well as  (\ref{integrals}), the integrals can be evaluated and we find
that the terms depending on Kronecker delta functions cancel out in
the final result after carrying out the $d^{3}k$ integration. This may
be seen by noticing that, when $\mu\nu\lambda$ are all spatial
indices, we can write (\ref{threepoint1}) in the form
\begin{eqnarray}
\Gamma_{ijl}^{\rm static} & = & -2ie^{3}T \int
\frac{d^{3}k}{(2\pi)^{3}}\,(\sin \tilde{p}_{1}\cdot k + \sin
\tilde{p}_{2}\cdot k + \sin \tilde{p}_{3}\cdot
k)\frac{\partial^{3}}{\partial k_{i}\partial k_{j}\partial k_{l}}
  \sum_{n} \log [(2\pi nT)^{2} + k^{2}]\nonumber\\
\noalign{\vskip 4pt}%
 & = & - 2ie^{3}T \tilde{p}_{1,i}\tilde{p}_{1,j}\tilde{p}_{1,l} \int
\frac{d^{3}k}{(2\pi)^{3}}\,\cos\tilde{p}_{1}\cdot k \sum_{n} \log
[(2\pi nT)^{2} + k^{2}] + {\rm two\ similar\ terms}\label{delta}
\end{eqnarray}
which shows that only terms involving triple products of the same
momentum are present in the final result for $\Gamma_{ijl}^{\rm static}$.

The nontrivial  components of
the three point amplitude, in the static limit, at leading order, then,
take the forms
\begin{eqnarray}
\Gamma_{00i}^{\rm static}(p_{1},p_{2},p_{3}) & = &
ie\left[\tilde{p}_{1,\,i}\,\Pi_{00}^{\rm static} (p_{1}) + {\rm
    cyclic}\right]\nonumber\\
\noalign{\vskip 4pt}%
\Gamma_{ijk}^{\rm static} (p_{1},p_{2},p_{3}) & = &
ie\left[\tilde{p}_{1,\,k}\,\Pi_{ij}^{\rm
    static} (p_{1}) + {\rm cyclic}\right]\label{threepointfinal}
\end{eqnarray}
It now follows from (\ref{threepointfinal}) that
\begin{eqnarray}
p_{3,\,i}\Gamma_{00i}^{\rm static} (p_{1},p_{2},p_{3}) & = &
ie\left[p_{3}\cdot\tilde{p}_{1}\, \Pi_{00}^{\rm static} (p_{1}) +
  p_{3}\cdot\tilde{p}_{2}\, \Pi_{00}^{\rm static}
  (p_{2})\right]\nonumber\\
\noalign{\vskip 4pt}%
 & \approx & - 2ie\sin \left(\frac{\tilde{p}_{1}\cdot p_{2}}{2}\right)
\left(\Pi_{00}^{\rm
  static} (p_{1}) - \Pi_{00}^{\rm static} (p_{2})\right)\nonumber\\
\noalign{\vskip 4pt}%
p_{3,\,k}\Gamma_{ijk}^{\rm static} (p_{1},p_{2},p_{3}) & = &
ie\left[p_{3}\cdot \tilde{p}_{1}\,\Pi_{ij}^{\rm static} (p_{1}) + p_{3}\cdot
  \tilde{p}_{2}\,\Pi_{ij}^{\rm static} (p_{2})\right]\nonumber\\
\noalign{\vskip 4pt}%
 & \approx & - 2ie\sin \left(\frac{\tilde{p}_{1}\cdot
  p_{2}}{2}\right)\left[\Pi_{ij}^{\rm static} (p_{1}) - \Pi_{ij}^{\rm
    static} (p_{2})\right]\label{wardidentity}
\end{eqnarray}
where we have used the conservation of momentum in the intermediate
steps as well as (\ref{approx}) to write
\begin{equation}
\tilde{p}_{1}\cdot p_{2} \approx 2\sin \left(\frac{\tilde{p}_{1}\cdot
  p_{2}}{2}\right)\label{approx1}
\end{equation}
This shows that the three point functions indeed satisfy simple Ward
identities and that all the nontrivial components of the three point
amplitude can, in fact, be determined from a knowledge of the
self-energy. 

\begin{figure}[h!]
% The following should be used when creating pdf directly out of
% latex, using the command line pdflatex
%    \includegraphics[scale=0.8]{four_photon.pdf}
\includegraphics[scale=0.8]{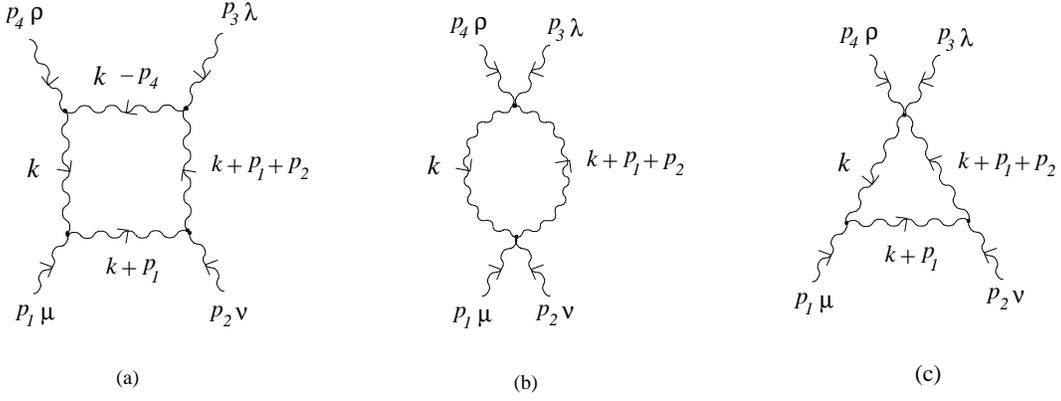}
\caption{Typical one-loop diagrams for the four point photon amplitude in
  non-commutative QED.}\label{fig4}
  \end{figure}

The general procedure outlined above can be used to evaluate the four
point amplitude (see figure \ref{fig4})
in the leading order of the derivative expansion. In the static limit,
this amplitude has the  form 
\begin{eqnarray}
\Gamma_{\mu\nu\lambda\rho}^{\rm static} (p_{1},p_{2},p_{3},p_{4}) & =
& 32 e^{4} \int 
\frac{d^{3}k}{(2\pi)^{3}}\,\sin\left(\frac{\tilde{p}_{1}\cdot
  k}{2}\right)\sin\left(\frac{\tilde{p}_{2}\cdot
  k}{2}\right)\sin\left(\frac{\tilde{p}_{3}\cdot
  k}{2}\right)\sin\left(\frac{\tilde{p}_{4}\cdot
  k}{2}\right)\nonumber\\
\noalign{\vskip 4pt}%
 &  & \!\!\times
T\sum_{n}\left[\frac{24k_{\mu}k_{\nu}k_{\lambda}k_{\rho}}{((2\pi
    nT)^{2} + k^{2})^{4}} +
    \left(-\frac{4\delta_{\mu\nu}k_{\lambda}k_{\rho}}{((2\pi nT)^{2} +
      k^{2})^{3}} + \frac{\delta_{\mu\nu}\delta_{\lambda\rho}}{((2\pi
        nT)^{2} + k^{2})^{2}} + {\rm
      permutations}\!\right)\!\right]\label{4point}
\end{eqnarray}
As in the case of the three point function, this expression
 simplifies, in practice, upon using the trigonometric identity 
\begin{eqnarray}
8\sin\left(\frac{\tilde{p}_{1}\cdot
  k}{2}\right)\sin\left(\frac{\tilde{p}_{2}\cdot
  k}{2}\right)\sin\left(\frac{\tilde{p}_{3}\cdot
  k}{2}\right)\sin\left(\frac{\tilde{p}_{4}\cdot k}{2}\right) & = &
  C(p_{1},k)  + C(p_{2},k) + C(p_{3},k) + C(p_{4})\nonumber\\
\noalign{\vskip 4pt}%
 &   &  -
  C(p_{1}+p_{4},k) - C(p_{2}+p_{4},k) - C(p_{3}+p_{4},k)\label{identity}
\end{eqnarray}
where we have defined
\begin{equation}
C(p,k) = 1 - \cos \tilde{p}\cdot k\label{cosine}
\end{equation}
For the spatial components, the integrand in (\ref{4point}) can be
written in  a similar form as in
(\ref{delta}), so that no Kronecker delta functions appear in the
final result when the $d^{3}k$ integration is carried out. Then, using
(\ref{identity}), we obtain
\begin{eqnarray}
\Gamma_{ijkl}^{\rm static} (p_{1},p_{2},p_{3},p_{4}) & = &
e^4\left[
f(\tilde{p}_{1}) \tilde{p}_{1,
  i}\tilde{p}_{1,j}\tilde{p}_{1,k}\tilde{p}_{1,l} + f(\tilde{p}_{2})
\tilde{p}_{2,i}\tilde{p}_{2,j}\tilde{p}_{2,k}\tilde{p}_{2,l} +
\cdots\right.\nonumber\\
\noalign{\vskip 4pt}%
 &  & -\left.
f(\tilde{p}_{1}+\tilde{p}_{4})(\tilde{p}_{1}+\tilde{p}_{4})_{i}
(\tilde{p}_{1}+\tilde{p}_{4})_{j}(\tilde{p}_{1}+\tilde{p}_{4})_{k}
(\tilde{p}_{1}+\tilde{p}_{4})_{l} - \cdots\right]\label{f}
\end{eqnarray}
where
\begin{equation}
f (\tilde{p}) = \frac{\tilde{\Pi}_{\rm T}^{\rm static}
  (\tilde{p})}{e^2\,\tilde{p}^{2}}\label{53}
\end{equation}
and $\tilde{\Pi}_{\rm T}^{\rm static}$ is given in
(\ref{formfactors1}). Using (\ref{threepointfinal}), this can be
written in terms of the three point amplitudes as
\begin{eqnarray}
\Gamma_{ijkl}^{\rm static} (p_{1},p_{2},p_{3},p_{4}) & = &
ie\left[\tilde{p}_{1,l} \Gamma_{ijk}^{\rm static}
  (p_{1}+p_{4},p_{2},p_{3}) + \tilde{p}_{2,l} \Gamma_{ijk}^{\rm
    static} (p_{1},p_{2}+p_{4},p_{3})\right.\nonumber\\
\noalign{\vskip 4pt}%
 &  & \quad \left. + \tilde{p}_{3,l} \Gamma_{ijk}^{\rm static}
(p_{1},p_{2},p_{3}+p_{4}) + \cdots\right]
\end{eqnarray}
where $\cdots$ represent terms needed to Bose symmetrize the amplitude.
It is easy to see that this form is consistent with the static Ward
identity    
\begin{eqnarray}
p_{4,l} \Gamma_{ijkl}^{\rm static} (p_{1},p_{2},p_{3},p_{4}) & = & ie\left[
(\tilde{p}_{1}\cdot 
p_{4}) \Gamma_{ijk}^{\rm static} (p_{1}+p_{4},p_{2},p_{3}) +
(\tilde{p}_{2}\cdot p_{4}) \Gamma_{ijk}^{\rm static}
(p_{1},p_{2}+p_{4},p_{3})\right.\nonumber\\
\noalign{\vskip 4pt}%
 &  & \qquad\left. + (\tilde{p}_{3}\cdot p_{4})
\Gamma_{ijk}^{\rm static} (p_{1},p_{2},p_{3}+p_{4})\right]
\end{eqnarray}

It is clear from these discussions of the static three and four point
amplitudes that the components, where not all the indices are
temporal, satisfy simple Ward identities, which follows from invariance
under a static gauge transformation. Such components can, therefore,
be recursively related.  (The reason why such simple Ward identities
hold in our case may be understood by noting that the contributions of
the ghost particles, to this order, cancel out in the BRST
identities.) The component of the four point amplitude with all
temporal  indices, $\Gamma_{0000}^{\rm static}$, on
the other hand, is not constrained in the static limit and, therefore,
cannot be related to lower order amplitudes. However, this component
can be evaluated from (\ref{4point}) and it can be seen, after some
algebra, that $\Gamma_{0000}^{\rm static}$ does not vanish. As a
result, this can be taken as a new perturbative input in determining
the complete static effective action. In fact, there will be a new
perturbative input at every even order in
perturbation, whenever the component of the amplitude with all
temporal indices does not vanish.

\section{The effective generating functional}

The analysis of the previous section shows that all the nontrivial
components of the three point function can be determined from a
knowledge of  the
self-energy. However, at the level of the four point function, we also
saw that we need to determine $\Gamma_{0000}$ independently since it is
invariant under static gauge transformations. This component of the
four point amplitude, on the other hand, would be essential in
determining all the components of the five point amplitude. In fact,
at every  even
order of the amplitudes, we expect new independent structures that
cannot be determined from a knowledge of the lower order
amplitudes. Therefore, it would be impossible to obtain a closed form
expression for the complete effective action from a knowledge of the
amplitudes to a given order. On the other hand, as we have seen, the
components  of the amplitudes with spatial indices
only are related recursively, through Ward identities, to lower order
amplitudes. Therefore, we can try to determine that part of the
effective action which depends only on $A_{i}$.

Let $\Gamma[A_{i}]$ represent the part of the effective action at high
temperature that
depends only on the spatial components of the gauge field. Then,
invariance under an infinitesimal  static gauge transformation, leads
to  the Ward identity
\begin{equation}
\frac{\delta\Gamma[A_{k}]}{\delta \omega (x)} = \int dy\,\frac{\delta
  A_{i}(y)}{\delta \omega (x)} \frac{\delta\Gamma [A_{k}]}{\delta A_{i}(y)}
  = D_{i}\,\frac{\delta \Gamma[A_{k}]}{\delta A_{i}(x)} = 0\label{ward}
\end{equation}
where $\omega(x)$ represents the infinitesimal gauge transformation
parameter depending only on the spatial coordinates. Equation
(\ref{ward})  is simply a statement of the covariant conservation of
current. Furthermore, under the approximation that we are using (see
(\ref{approx})), the covariant derivative, in the adjoint
representation,  takes the form
\begin{equation}
D_{i} = \partial_{i} + e (\partial_{j}
A_{i})\,\tilde{\partial}_{j}\label{covariantderivative}
\end{equation}
With this, the current conservation, (\ref{ward}), takes the form
\begin{eqnarray}
\partial_{i}\,\frac{\delta\Gamma[A_{k}]}{\delta A_{i}} + e
(\partial_{j}
A_{i})\,\tilde{\partial}_{j}\,\frac{\delta\Gamma[A_{k}]}{\delta A_{i}}
& = & 0\nonumber\\
\noalign{\vskip 4pt}%
{\rm or,}\quad \partial_{i}\left(\frac{\delta\Gamma[A_{k}]}{\delta
  A_{i}} + e
A_{j}\tilde{\partial}_{i}\,\frac{\delta\Gamma[A_{k}]}{\delta
  A_{j}}\right) & = & 0\label{ward1}
\end{eqnarray}
This determines that the quantity in the parenthesis vanishes up to a
term that is transverse, namely,
\begin{equation}
\left(\delta_{ij} + e
A_{j}\tilde{\partial}_{i}\right)\frac{\delta\Gamma[A_{k}]}{\delta
  A_{j}} = X_{i}^{\rm T}\label{ward2}
\end{equation}
such that
\begin{equation}
\partial_{i}X_{i}^{\rm T} = 0
\end{equation}

By taking the functional derivative of (\ref{ward2}) with respect to
$A_{j}$ and setting all the fields to zero, it can be easily
determined that, to lowest order
\begin{equation}
X_{i}^{{\rm T}\,{\rm (lowest)}} = \Pi_{ij}^{\rm static}A_{j}\label{x}
\end{equation}
It is clear that $X_{i}^{\rm T}$ will contain higher order terms in
the fields as
well. However, it can be seen by taking higher order functional
derivatives  of (\ref{ward2}) that the role of the higher order terms
in $X_{i}^{\rm T}$ is to Bose symmetrize the higher point
amplitude.  Thus, keeping this Bose symmetrization
in mind, we can neglect the contributions involving higher order terms
in the fields in 
$X_{i}^{\rm T}$. In such a case, we can solve for the current from
(\ref{ward2}) and obtain
\begin{equation}
\frac{\delta\Gamma[A_{k}]}{\delta A_{i}} = \left(\delta_{ij} + e
A_{j}\tilde{\partial}_{i}\right)^{-1}\,X_{j}^{{\rm T}\,{\rm
    (lowest)}}\label{soln}
\end{equation}

The quantity in the parenthesis on the right hand side in (\ref{soln})
is an operator and hence does not have a unique left-right
inverse. However, the one that is relevant, for the solution, is the
right  inverse which can be determined to be
\begin{equation}
\left(\delta_{ij} + eA_{j}\tilde{\partial}_{i}\right)^{-1} =
\delta_{ij} - e A_{j} \tilde{\partial}_{i} + e^{2}
A_{k}\tilde{\partial}_{i}A_{j}\tilde{\partial}_{k} - e^{3}
A_{k}\tilde{\partial}_{i}A_{l}\tilde{\partial}_{k}A_{j}\tilde{\partial}_{l} 
+ \cdots\label{inverse}
\end{equation}
Furthermore, we recognize from the definition of the covariant
derivative (\ref{covariantderivative}) that
\begin{equation}
\partial_{j}\left(\delta_{ji} + e A_{i}\tilde{\partial}_{j}\right) =
D_{i}
\end{equation}
so that we can also write
\begin{equation}
\left(\delta_{ij} + e A_{j}\tilde{\partial}_{i}\right)^{-1} =
D_{j}^{-1}\partial_{i}\label{covariant1}
\end{equation}
Using (\ref{covariant1}), we can determine the current in (\ref{soln})
to be
\begin{equation}
j_{i}[A_{k}] = \frac{\delta\Gamma[A_{k}]}{\delta A_{i}} =
D_{j}^{-1}\partial_{i} \Pi_{jk}^{\rm static} A_{k}\label{current}
\end{equation}
We note that this current manifestly satisfies covariant conservation
since the self-energy is transverse. Furthermore, this closed form
expression  for the current can be explicitly checked to lead to the
correct amplitudes, under Bose symmetrization.

The current is all we need for the generation of any
amplitude. However, it will also be nice to determine the static effective
action in a closed form. That involves functionally integrating the
current  which
appears to be highly nontrivial. Nevertheless, we can obtain the
effective action as explained in the next section.

\section{Discussion}

Here we present a closed-form effective action for the static amplitudes
(with spatial tensor indices) valid in the region (as in (39))
\be |p_a| \ll T,~~ |p_a| \ll |{\theta} |^{-1/2}, \label{jct0}\ee
where $a=1,2,\dots$ runs over the external momenta.
In this region, we expect the internal momentum $k$ to be of the
order given in  (\ref{hardthermalloop}).

Let us first define
\be C(p,A)= \int d^4x \exp[-ip\cdot x+ie{\tilde p}\cdot A(x)]. \label{jct1}\ee
This is a function of an auxiliary 4-momentum $p$
and a functional of (the spatial components of) $A$.  
We will identify $p$ with the linear
combinations of external momenta, as in (\ref{f}).
In the region (\ref{jct0}), the general gauge-transformation 
(\ref{gauge transformation}) may be approximated as
\be{\delta} A_i(x)=[{\partial}_i+ie({\tilde\partial}_jA_i(x))
{\partial}_j]{\omega}(x). \label{jct2}\ee
$C$, defined in (\ref{jct1}), is invariant under (\ref{jct2}).
To prove this, we note that
\[{\delta}\left({\tilde p}\cdot A\right)=
[{\tilde p}\cdot {\partial}+e({\tilde\partial}_j{\tilde p}\cdot A(x))
{\partial}_j]{\omega}(x),\]
\newpage
\[{\delta}[\exp(ie{\tilde p}\cdot A(x)]=i\exp(ie{\tilde p}\cdot A)
[{\tilde p}\cdot {\partial}+e({\tilde\partial}_j
{\tilde p}\cdot A(x)){\partial}_j]{\omega}(x)\]
\be=i\exp(ie{\tilde p}\cdot A){\tilde p}\cdot {\partial}{\omega}
-{\tilde\partial}_j[\exp(ie{\tilde p}\cdot A){\partial}_j{\omega}], 
\label{jct3}\ee
where we have used $\tilde{\partial}\cdot \partial = 0$.
Substituting (\ref{jct3}) into (\ref{jct1}) and integrating 
by parts (so that ${\tilde\partial}_j$ differentiates the
$e^{-ip\cdot x}$) we obtain
\be{\delta} C=i\int d^4x e^{-ip\cdot x} 
\exp(ie{\tilde p}\cdot A)[{\tilde p}\cdot {\partial}{\omega}
-{\tilde p}\cdot {\partial}{\omega}]=0. \label{jct4}\ee

Now we can construct the effective action in terms of $C$:
\be \Gamma={1\over 2\times (2\pi)^8} 
\int d^4p f({\tilde p}) C(p, A)C(-p, A)={1\over 2
\times (2\pi)^{8}}\int d^4p
f({\tilde p})|C(p,A)|^2, \label{jct5}\ee
where  $f$ is defined by (\ref{53}).

That $\Gamma$ in (\ref{jct5}) is the correct effective action follows because
it trivially agrees with  (\ref{components}) and (\ref{formfactors1})
to order $e^2$,  it is gauge-invariant, and it gives the functional
dependence on the ${\tilde p}_a$ ($a=1,2,3,\dots\nu$ for the 
$\nu$-point function) typified by (52).
We have verified explicitly that $\Gamma$ gives the 3- and 4-point 
functions correctly.

It is much more difficult to find an effective action, not assuming both
inequalities in (\ref{jct0}), but just 
\be |p_a|\ll T. \label{jct7}\ee
In this case we must use the exact gauge transformation 
(\ref{gauge transformation}), not just the approximate one in (\ref{jct2}).
But we note that $C$ in (\ref{jct1}) does have a generalization
which is gauge-invariant under the exact gauge-transformation 
(\ref{gauge transformation}). This generalization is 
\be W(p,A)=\int {\rm d}^4 x \exp{(-i p\cdot x)}\star
P\exp{\left[ie\int_0^1 {\rm d}\xi\, \tilde p\cdot 
A(x+\xi \tilde p)\right]},
\label{jct7a}\ee
where $P$ denotes path ordering on the manifold characterized by the
star product (\ref{starproduct}). $W(p,A)$ represents the Fourier
transform of a gauge invariant open Wilson line, extending along a
straight path from $x$ to 
$x+\tilde p$ \cite{VanRaamsdonk:2001jd,Armoni:2001uw}.
Note that, if (\ref{jct0}) is assumed, $W$ reduces just to $C$.

However, the thermal effective action (when (\ref{jct0}) is not assumed)
is not obtained just by replacing $C$ by $W$ in (\ref{jct5}). 
The reason is that  the internal photon momentum $k$ is expected
to be of order $1/(|\theta||p_a|)$ and therefore 
(without (\ref{jct0})) we cannot make the hard thermal loop 
approximation of neglecting $|p_a|$ compared to $|k|$. 
The amplitudes are then much more complicated, and we cannot
expect them to be expressible in terms of a single function $f$
as in (\ref{jct5}).

\noindent{\bf Acknowledgment:}

We would like to thank the referee for helpful comments.
This work was supported in part by US DOE Grant number DE-FG
02-91ER40685, by CNPq and FAPESP, Brasil.


\newpage

\begin{thebibliography}{10}

%1
\bibitem{kapusta:book89lebellac:book96das:book97}
J.~I. Kapusta, {\em Finite Temperature Field Theory} (Cambridge
University
  Press, Cambridge, England, 1989);\\
%
%\bibitem{lebellac:book96}
M.~L. Bellac, {\em Thermal Field Theory} (Cambridge University
Press,
  Cambridge, England, 1996); \\
%
%\bibitem{das:book97}
A. Das, {\em Finite Temperature Field Theory} (World Scientific,
NY, 1997).
%5
\bibitem{weldon}
H. A. Weldon, Phys. Rev. {\bf D47}, 594 (1993), P. F. Bedaque and
A. Das, Phys. Rev. {\bf D47}, 601 (1993), A. Das and M. Hott,
Phys. Rev. {\bf D53}, 2252 (1996).

\bibitem{Braaten:1990it}
E. Braaten and R.~D. Pisarski, 
%``Calculation of the gluon damping rate in hot QCD,'' 
%Phys. Rev. {\bf D42,} 2156--2160 (1990);
%
%\bibitem{braaten:1992gm}
%E. Braaten and R.~D. Pisarski, 
%``Simple effective Lagrangian for hard thermal loops,'' 
%Phys. Rev. {\bf D45,} 1827--1830 (1992);
%
%\bibitem{braaten:1990mz}
%E. Braaten and R.~D. Pisarski, 
%``SOFT AMPLITUDES IN HOT GAUGE THEORIES: A GENERAL ANALYSIS,'' 
Nucl. Phys. {\bf B337,} 569 (1990);
%
%\bibitem{braaten:1990az}
%E. Braaten and R.~D. Pisarski, 
%``DEDUCING HARD THERMAL LOOPS FROM WARD IDENTITIES,'' 
Nucl. Phys. {\bf B339,} 310 (1990).
%2
\bibitem{frenkel:1991ts}
J. Frenkel and J.~C. Taylor, 
%``Hard thermal QCD, forward scattering and effective actions,'' 
Nucl. Phys. {\bf B334,} 199 (1990) ;Nucl. Phys. {\bf B374,} 156 (1992).
%6
\bibitem{Seiberg:1999vs}
N. Seiberg and E. Witten,
%``String theory and noncommutative geometry,'' 
JHEP {\bf 09,} 032 (1999).
%13
\bibitem{Fischler:2000fv}
W. Fischler, J. Gomis, E. Gorbatov, A. Kashani-Poor, S. Paban, and P. Pouliot,
%``Evidence for winding states in noncommutative quantum field
% theory,'' 
JHEP {\bf 05,} 024 (2000);
W. Fischler, E. Gorbatov, A. Kashani-Poor, R. McNees, S. Paban, P. Pouliot, 
JHEP {\bf 06,} 032 (2000).

\bibitem{hayakawa}
M. Hayakawa, Phys. Lett. {\bf B478}, 394 (2000).
%14
\bibitem{Arcioni:1999hw}
G. Arcioni and M.~A. Vazquez-Mozo, 
%``Thermal effects in perturbative noncommutative gauge theories,'' 
JHEP {\bf 01,} 028 (2000).
%15 
\bibitem{Landsteiner:2000bw}
K. Landsteiner, E. Lopez, and M.~H.~G. Tytgat, 
%``Excitations in hot noncommutative theories,''
JHEP {\bf 09,} 027 (2000);
%
%\bibitem{landsteiner:2001ky}
%K. Landsteiner, E. Lopez, and M.~H.~G. Tytgat, 
%``Instability of noncommutative SYM theories at finite
% temperature,''
JHEP {\bf 06,} 055 (2001).
%16
\bibitem{Szabo:2001kg}
R.~J. Szabo, 
%``Quantum Field Theory on noncommutative spaces,''   
hep-th/0109162 (2001)
%17
\bibitem{Douglas:2001ba}
M.~R. Douglas and N.~A. Nekrasov, 
%``Noncommutative field theory,''
Rev. Mod. Phys. {\bf 73,} 977 (2002).
%19 
\bibitem{Chu:2001fe}
A.~Armoni, R.~Minasian and S.~Theisen,
%``On non-commutative N = 2 super Yang-Mills,''
Phys.\ Lett.\ B {\bf 513}, 406 (2001);
C.-S. Chu, V.~V. Khoze, and G. Travaglini, 
%``Dynamical breaking of supersymmetry in 
% noncommutative gauge theories,'' 
Phys. Lett. {\bf B513,} 200 (2001);
%
%\bibitem{Chu:2001kq}
C.-S. Chu, V.~V. Khoze, and G. Travaglini,
%``Noncommutativity and model building,''   
hep-th/0112139 (2001).
%20
\bibitem{VanRaamsdonk:2001jd}
M. Van~Raamsdonk,
%``The meaning of infrared singularities in noncommutative
% gauge theories,'' 
JHEP {\bf 11,} 006 (2001).
%21
%\cite{Armoni:2001uw}
\bibitem{Armoni:2001uw}
A.~Armoni and E.~Lopez,
%``UV/IR mixing via closed strings and tachyonic instabilities,''
Nucl.\ Phys.\ B {\bf 632}, 240 (2002).
%[arXiv:hep-th/0110113].
%%CITATION = HEP-TH 0110113;%%
%28
\bibitem{Bonora:2000ga}
A.~Armoni,
%``Comments on perturbative dynamics of non-commutative Yang-Mills theory,''
Nucl.\ Phys.\ B {\bf 593}, 229 (2001);
L. Bonora and M. Salizzoni,
%``Renormalization of noncommutative U(N) gauge theories,''
Phys. Lett. {\bf B504,} 80 (2001).
%18
%\cite{Brandt:2002rw}
\bibitem{Brandt:2002rw}
F.~T.~Brandt, A.~Das, J.~Frenkel, D.~G.~C.~McKeon and J.~C.~Taylor,
%``Transport equation and hard thermal loops in noncommutative
%Yang-Mills  theory,'' 
Phys. Rev. {\bf D66}, 045011 (2002).
%%CITATION = HEP-TH 0204192;%%
%22 
\bibitem{gomis}
See, for example, J. Gomis and T. Mehen, Nuc. Phys. {\bf B591}, 265
(2000), O. Aharony, J. Gomis and T. Mehen, JHEP {\bf 09}, 023 (2000).

\bibitem{sean} M. Chaichian, M. M. Sheikh-Jabbari and A. Tureanu,
  Phys. Rev. Lett. {\bf 86}, 2716 (2001); S. M. Carroll, J. A. Harvey,
  V. A. Kostelecky, C. D. Lane and T. Okamoto, Phys. Rev. Lett. {\bf 87},
  141601 (2001).
 
%\cite{Brandt:2002aa}
\bibitem{Brandt:2002aa}
F.~T.~Brandt, J.~Frenkel and D.~G.~C.~McKeon,
%``Hard thermal effects in noncommutative U(N) Yang-Mills theory,''
Phys.\ Rev.\ D {\bf 65}, 125029 (2002).
%%CITATION = HEP-TH 0202202;%%
%18p

\bibitem{Sheikh-Jabbari:1999vm}
M.~M. Sheikh-Jabbari, 
%``Open strings in a B-field background as electric dipoles,''
Phys. Rev. Lett. {\bf 84}, 5265 (2000).
%24
\bibitem{fernando}
F. T. Brandt, A. Das and J. Frenkel, Phys. Rev. {\bf D65}, 085017
(2002).

\bibitem{gradshteyn} I. S. Gradshteyn and M. Ryzhik, ``Table of
  Integrals, Series and Products'' (Academic, New York, 1980).

\bibitem{brandt:1993mj}
F.~T. Brandt, J. Frenkel, J.~C. Taylor, and S.~M.~H. Wong,
% ``Effective actions
% for Braaten-Pisarski resummation,''
Can. J. Phys. {\bf 71,} 219 (1993).
%3
\bibitem{brandt:1993dkbrandt:1997se}
F.~T. Brandt and J. Frenkel, 
%``The Three graviton vertex function in thermal quantum gravity,'' 
Phys. Rev. {\bf D47,} 4688 (1993);
%4
%\bibitem{brandt:1997se}
%F.~T. Brandt and J. Frenkel, 
%``Generalized forward scattering amplitudes in QCD at high temperature,'' 
Phys. Rev. {\bf D56,} 2453 (1997).

\end{thebibliography}
\end{document}